# High-Dynamic Range Broadband Terahertz Time-Domain Spectrometer Based on Organic Crystal MNA

**S. Mansourzadeh[1, *], T. Vogel[1], A. Omar[1], M. F. Biggs[2], E. S.-H. Ho[2], C. Hoberg[3], D. J. Michaelis[2], M. Havenith[3], J. A. Johnson[2], C. J. Saraceno[1]**

[1]*Photonics and Ultrafast Laser Science (PULS), Ruhr-University Bochum, Germany*
[2]*Department of Chemistry and Biochemistry, Brigham Young University, Provo, UT, USA*
[3]*Faculty of Chemistry and Biochemistry, Ruhr-University Bochum, Germany*

*\*mansourzadeh.samira@ruhr-uni-bochum.de*

**Abstract:** We present a high power and broadband THz-TDS setup utilizing the nonlinear organic crystal MNA both as emitter and detector. The THz source is based on optical rectification of near infra-red laser pulses at a central wavelength of 1036 nm from a commercial, high-power Yb-based laser system and reaches a high THz average power of 11 mW at a repetition rate of 100 kHz and a broad and smooth bandwidth of more than 9 THz. The conversion efficiency is high (0.13%) in spite of the high excitation average power of 8 W. We validate the high dynamic range and reliability of the source for applications in linear spectroscopy by measuring the broadband THz properties of $\chi^{(2)}$ nonlinear crystals up to 8 THz. This new high-repetition rate source is very promising for ultra-broadband THz spectroscopy at high dynamic range and/or reduced measurement times.

## 1. Introduction

In recent years, laser-driven pulsed terahertz (THz) sources have become widely popular in many fundamental and applied fields of science and technology. Many of the useful applications of these sources stem from the ultra-broad bandwidths generated, that can be detected with high sensitivity using coherent THz time-domain spectroscopy (THz-TDS). For most *linear* spectroscopy applications, the crucial metric to optimize the performance of THz-TDS is the dynamic range (DR) at a given desired frequency (or over a given bandwidth), as defined in [1]. The overall DR of THz-TDS measurement can be improved either by multiplying the single THz pulse strength (thus improving single-shot dynamic range) or via averaging over many pulses (reducing the noise floor, provided the laser does not suffer from strong shot-to-shot fluctuations), or ideally, a combination of both. In fact, the right metric to scale DR is average power, whereas a high repetition rate is usually desired as well for reduced measurement times, to enable fast detection methods and further averaging. So far, most linear spectroscopy experiments use laser systems with high repetition rates but low pulse strength, typically based on photoconductive emitters. These systems perform impressively in peak dynamic range [2], but usually only allow reliable parameter extraction up to 4 THz due to a strong rollover in dynamic range at high frequencies. Furthermore, they critically rely on ultra-low noise laser systems, which often restrict average power scaling possibilities.

Making use of novel THz-TDS with higher pump pulse strength and high repetition rate is an attractive alternative that is underexplored for achieving high DR. This offers the advantage of emitters with potentially much broader bandwidths, typically associated with low DR. In fact, the challenges related to slow acquisition times become significantly more difficult to tackle when higher THz frequencies are desired, i.e. significantly above 5 THz. Sources capable of such broad bandwidths often require high driving peak power, which in turn comes with high pulse energies and low repetition rates [3,4]. In contradiction with low repetition rates, detection methods covering ultra-broad bandwidths are typically less sensitive, e.g. the THz detection based on third-order nonlinear susceptibility [5], thus the requirement for high repetition rates is even more critical to reach high DR. Many fields now require broadband and sensitive THz-TDS, such as observation of ultrafast electronic phase change in semiconductors [6], broadband characterization of emerging materials such as solar absorbers or metal oxide-based photoelectrodes [7,8] or applications in chemistry and biology [8–11]. In this context, increasing the repetition rate of ultra-broadband, high-DR sources over frequencies > 5 THz remains a crucial goal in THz technology.

Several approaches have been explored successfully in recent years to increase the average power of broadband THz schemes. Using inorganic crystals such as lithium niobate, a record THz power of 643 mW with a 1.5 THz bandwidth at 40 kHz repetition rate has been achieved [12]. The two-color plasma scheme for ultra-broad bandwidths has reached 640 mW at a 500 kHz repetition rate, however, the use of such a highly specialized high-energy laser system complicates its application [13].

An alternative solution to scale THz average power and bandwidth while boosting detection sensitivity, with moderate driving pulse energy, is optical rectification (OR) and electro-optic sampling (EOS) in organic crystals. In the low repetition rate regime (<1 kHz), these crystals show promising results: a high energy of 0.9 mJ [14], a high electric field of 6 MV/cm and a high average power of 68 mW at a repetition rate of 100 Hz [15], and a conversion efficiency of 6% at the repetition rate of 20 Hz [16]. For higher repetition rates, thermal management becomes critical which can be circumvented via burst mode excitation [17]. In this way, 5.6 mW of average power has been demonstrated at 540 kHz repetition rate [18] and up to mW levels at MHz regime [19,20]. Among these efforts, MHz repetition rate sources are significantly more challenging to scale the conversion efficiency and produce rather moderate THz power and field, limiting many applications in spectroscopy. In this regard, repetition rates of hundreds of kHz seem to offer an attractive middle ground for high DR and efficient THz generation using well-established industrial lasers, and still enable fast acquisition schemes compatible with fast averaging. Besides their use in THz generation, organic crystals are also advantageous for THz detection. They enable the detection of the entire generated bandwidth, ensuring comprehensive measurement of THz signals [21].

Among various organic crystals validated in the literature, MNA (2-Amino-5- Nitrotoluene) was identified more than 40 years ago as a promising material for nonlinear applications due to its high molecular hyperpolarizability, favorable non-centrosymmetric crystal structure, and high nonlinear coefficient of 250 pm/V [22]. However, it has not been widely used to generate THz radiation due to the difficulties in synthesizing large crystal sizes. Only very recently it was possible to grow large single crystals which are suitable for THz generation which resulted in a conversion efficiency of 3% using a pump laser with a central wavelength of 1250 nm and a repetition rate of 1 kHz [23]. However, MNA has so far not been tested with 1030 nm wavelength nor been explored for high repetition rate.

In this paper, we demonstrate a THz-TDS based on OR and EOS detection using the organic crystal MNA at room temperature, driven by a commercial, high-power femtosecond laser system at 1036 nm central wavelength and 100 kHz repetition rate. The driving laser pulses are temporally compressed from 220 fs down to 50 fs using a home-built multi-pass cell (MPC). In optimized conditions, using 8 W of driving power on the crystal, we reach 11 mW of THz average power with a broad bandwidth extending up to 9 THz, at a peak DR of 55 dB, with a conversion efficiency of 0.13%. Remarkably, a DR > 40 dB is achieved over most of the detected bandwidth in 5 minutes of measurement time. In order to achieve this performance, we use MNA also as a broadband and sensitive detector, which is typically not possible to realize due to the poor quality of organic crystals. Our TDS offers an unprecedented combination of usable bandwidth, DR and average power for applications. We demonstrate the impact of the source for applications in broadband linear spectroscopy. More particularly, we use our source to characterize the THz properties of gallium phosphide (GaP) and 4-Nitro-4′-methylbenzylidene aniline (NMBA) over a bandwidth of 8 THz, exemplifying the high DR performance over the whole bandwidth available. The complex refractive indices are extracted using our recently reported, open-source software phoenix [24]. In addition, we examine the stability of the THz source by recording the amplitude of the THz trace in multiple days in row to demonstrate its high reliability for linear spectroscopy applications. Last but not least, our source also achieves high peak field of 212 kV/cm, making it a potentially viable candidate for nonlinear spectroscopy. We believe the TDS demonstrated – based on commercially available high-power laser technology – offers great promise for broadband THz applications. To the best of our knowledge this is the first report of a THz source based on an organic crystal MNA using 1036 nm high average power excitation in combination with MNA based detection which also includes the system reliability tests and linear spectroscopy experiments.

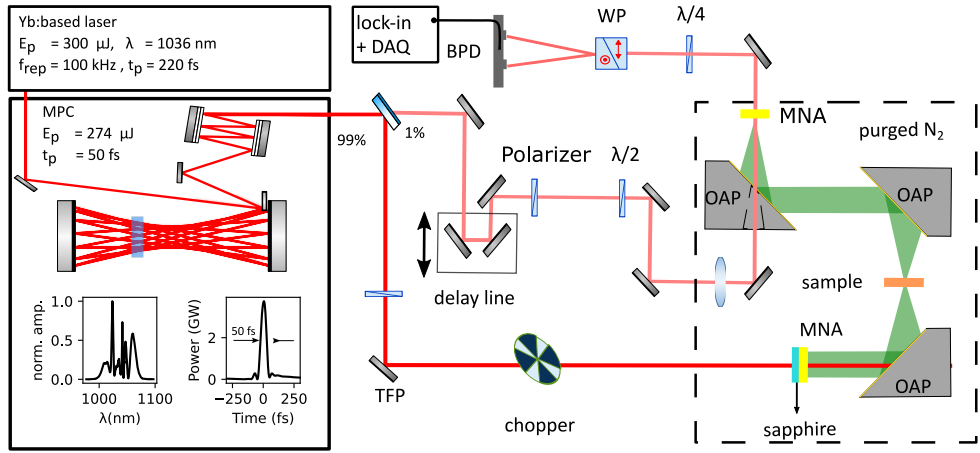

Fig. 1. Full experimental setup consisting of driving laser, MPC, and THz-TDS parts. The inset shows the spectrum and temporal profile of the pump laser retrieved from the FROG trace.
BPD: balanced photodetector, DAQ: data acquisition, WP: Wollaston prism.

## 2. Experimental setup

The experimental setup is shown in Fig. 1. The laser system is an industrial, Yb-based amplifier system (Light Conversion, CARBIDE) providing up to 400 µJ pulse energy, but only 300 µJ is used in the current experiment, as the rest is reserved for future pump-probe experiments. The pulses from the driving laser are temporally compressed from 220 fs down to 50 fs using a home-built, Herriott-type MPC spectral broadening stage and subsequent pulse compression with broadband chirped mirrors. The total pulse energy after the MPC is 274 µJ (average power of 27.4 W). The MPC is characterized using a frequency resolved optical gating (FROG) setup. The spectrum and temporal pulse profile retrieved from the FROG trace are shown in Fig. 1 (inset). After the MPC, the laser beam is split into two parts: 99% is used to generate THz radiation in one MNA crystal with a thickness of 730 µm and dimension of about 20 mm × 4 mm. The (010) MNA crystal is directly fused, without adhesive as described in [25], on a sapphire substrate, and mounted in a metal rotation mount for improved thermal dissipation. It is mounted in such a way, that the pump laser passes through the sapphire before reaching the MNA crystal to avoid affecting the THz pulse. The reflection of the near-infrared (NIR) pump from the sapphire substrate is considered (loss of about 7.5%) in the calculation of THz conversion efficiency. The collimated pump beam has a $1/e^2$ diameter of 3.6 mm at the position of the MNA. To further reduce the thermal load on MNA, an optical chopper with a duty cycle of 50% is placed before the crystal. The pump power is adjusted on the crystal using a combination of a thin film polarizer (TFP) and a waveplate ($\lambda/2$). With a set of three off-axis parabolic (OAP) mirrors, two THz foci are generated. All OAPs in the setup have a focal length of 101.6 mm and a diameter of 50.8 mm. In the second focus point, the THz and probe pulses (1% of the total NIR laser power) are spatially and temporally overlapped on the detection crystal to reconstruct the THz electric field in the EOS setup. A polarizer is used to define the polarization of the probe arm, followed by a half-wave plate to rotate the polarization of the probe beam according to both the THz beam polarization and the orientation of the detection crystal. To maximize detectable bandwidth and DR at higher frequencies, we use a second MNA crystal with a thickness of 650 μm as the detection crystal. This approach is feasible due to the high surface and interior quality of the MNA crystal, which does not scatter the NIR optical probe beam.

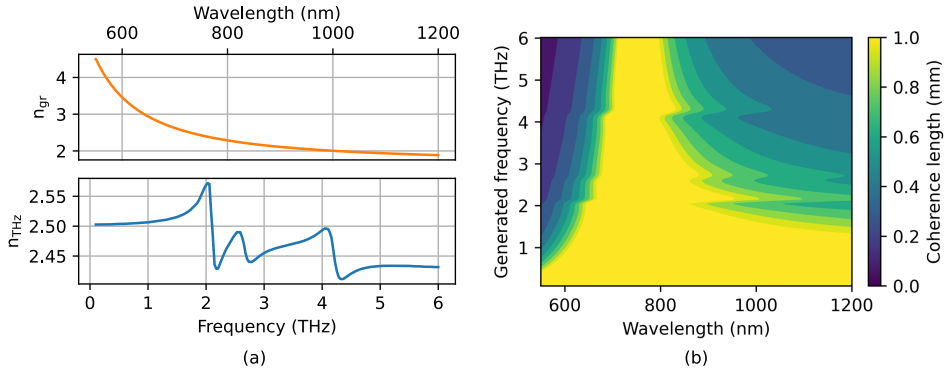

Fig. 2. (a) THz refractive index [23] and pump group refractive index for MNA [26]. The dashed line indicates the pump wavelength of 1030 nm for the corresponding group index. (b) calculated coherence length versus generated THz frequency and pump wavelength for MNA.

To filter out the residual laser radiation and the generated second-harmonic green light after MNA, a 2 mm Germanium (Ge) wafer (with 39.4% averaged THz transmission, accounting for Fresnel losses, internal absorption and echoes) is used. To measure the THz average power, the detection MNA crystal is replaced with a calibrated THz power meter (THz20, SLT GmbH). The modulation frequency of the chopper is reduced to 18 Hz, to reproduce the conditions at the calibration of the detector at the German metrology institute (Physikalisch-Technische Bundesanstalt, PTB). Additionally, to estimate the THz electric field, a sensitive THz camera (RIGI Camera, Swiss Terahertz) is used in the position of the second THz focus.

## 3. Experimental and simulation results and discussion

The experimental results are divided into two sections: the first section presents the development and optimization of the high-power, broadband, THz TDS. In the second section, this THz-TDS system is used to characterize the THz properties of two nonlinear materials over a broad frequency range to validate the performance in a TDS application.

### 3.1 Developing the high power, broadband THz source

To optimize the THz generation, we begin with velocity matching and calculating the coherence length in MNA, as these are crucial factors for efficient THz generation. In Fig. 2(a) the group refractive index curve of MNA in the pump wavelength regime [26] ($n_{gr}$, orange curve) and phase refractive index in the THz range up to 6 THz ($n_{THz}$, blue curve) [23] are shown. Fig. 2 (b) depicts the coherence length of MNA as a function of pump wavelength and THz frequency. The coherence length is a practical metric to evaluate the nonlinear crystal length that is tolerable before velocity mismatch between THz and pump waves becomes too large [17]. As it can be seen, the velocity matching for MNA with the most broadband operation is occurring at wavelength lower than 800 nm. However, large bandwidth can also be obtained for 1030 nm excitation, but slightly thinner crystals are required. However, there is a tradeoff between achieving higher THz power from thicker crystals and maintaining the generated frequency bandwidth. To find the optimum thickness of MNA at this pump wavelength, we simulated the generated spectrum by numerically modeling the THz generation process in MNA. For this simulation the coupled wave equations in 1+1D, i.e. considering the temporal dimension and the propagation direction are solved. It is conducted for THz pulse generation using OR in MNA under realistic phase matching and absorption conditions similar to [27]. The refractive index and absorption coefficient of MNA in the THz regime up to 6 THz are reported in [23]; the properties of MNA in proximity to the pump wavelength are taken from [26] and the nonlinear susceptibility is given in [22]. According to the simulation, the optimized thickness to get the broadest bandwidth which has the highest amplitude for higher frequency components is 690 μm. The available crystal thickness for our experiment is 730 μm which is close to the predicted optimum condition.

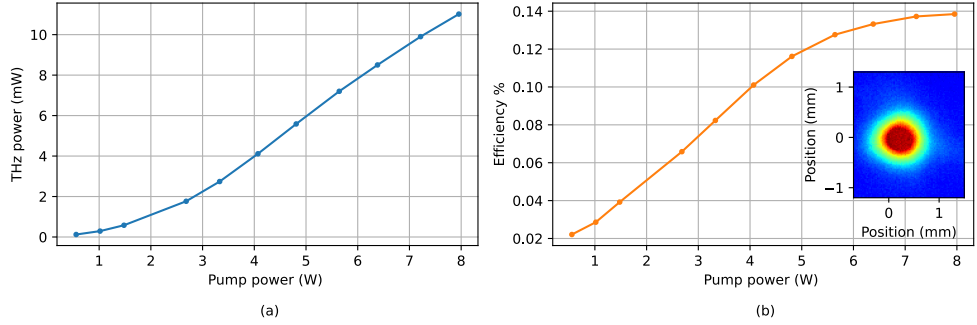

Fig. 3. THz power characterization using a power-meter. (a): THz average power vs. pump power. (b): corresponding conversion efficiency. The inset demonstrates the THz profile captured with the camera.

In the first measurement set, the power-meter is placed in the focus of the third OAP. Fig. 3 (a) shows the THz power, and Fig. 3 (b) indicates the THz conversion efficiency versus pump power. We can pump the crystal up to 8 W without any irreversible damage to the crystal. This maximum pump power corresponds to a fluence of 1.6 mJ/cm² and a peak intensity of 11.7 GW/cm². It should be mentioned that the maximum pump fluence applied to the crystal is comparable to the value reported in [23], which is 1.34 mJ/cm² at pump wavelength of 1250 nm and a repetition rate of 1 kHz. Using the pump power of 8 W, we achieve a maximum THz average power of 11 mW and a corresponding efficiency of 0.13%. It should be noted that the pump beam diameter of 3.6 mm is maximized to the size of the crystal to get the highest THz power for the maximum pump pulse energy, which can be applied without damaging the crystal. As shown in Fig. 3 (b), the efficiency curve reaches its peak at a pump power of 8 W (corresponding fluence of 1.6 mJ/cm²) which is lower than the maximum available pump power of 14 W after chopper (corresponding fluence of 2.25 mJ/cm²). To achieve higher THz power, it would be ideal for the efficiency peak to occur at a higher pump fluence. However, due to the limited crystal size, it was not possible to increase the pump spot diameter further on the crystal.

To detect the THz electric field, the power-meter is replaced with a detection crystal to characterize the electric field of the THz trace in the EOS setup. Our preliminary experiments, as we reported in [28], demonstrated the superior performance of MNA as a detection crystal compared to GaP. As a consequence, we use a 650 μm MNA crystal as the detection medium to sample the generated THz trace, utilizing approximately 300 mW of laser power. MNA, with space group Cc, for a (010) cut has specific zero and non-zero elements to the nonlinear susceptibility d-tensor and the electro-optic tensor. Using first principles calculations of the molecular hyperpolarizability of MNA together with the oriented gas model for the susceptibility [29,30], we can calculate simple values of these different tensor elements:

$$d_{ij} = \frac{1}{2} \begin{bmatrix} 44.3 & 3.2 & 4.4 & 0 & -16.4 & 0 \\ 0 & 0 & 0 & -1.2 & 0 & 3.2 \\ -16.4 & -1.2 & -0.23 & 0 & 4.4 & 0 \end{bmatrix} \quad r_{ij} = \begin{bmatrix} -2.3 & 0 & 0.84 \\ -0.16 & 0 & 0.06 \\ -0.23 & 0 & 0.01 \\ 0 & 0.06 & 0 \\ 0.84 & 0 & -0.23 \\ 0 & -0.16 & 0 \end{bmatrix}$$

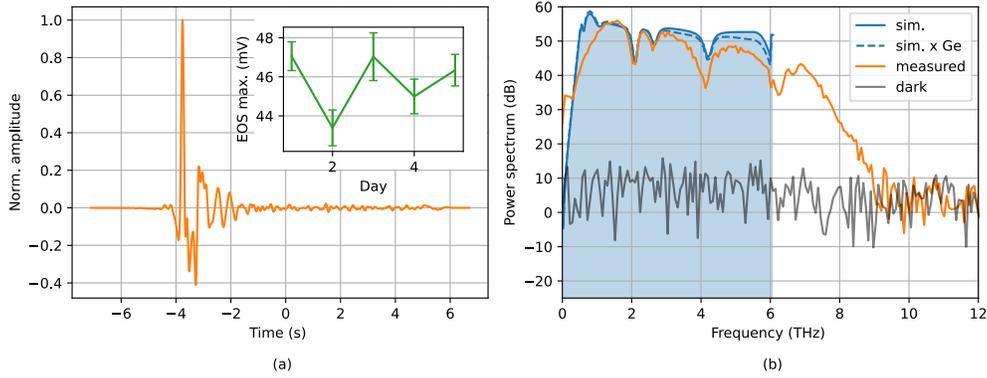

Fig. 4. EOS: (a) THz trace in time domain. The inset shows the maximum amplitude of the EOS traces recorded for 10 minutes for 5 days in row. (b) Corresponding spectrum in orange line and the dark measurement in grey line. Blue shaded area indicates the simulation results.

Both are reported in units of pm/V. For comparison of magnitudes, the $r_{41}$ value for (110) GaP is ~1 pm/V [31].

The EOS data is acquired using a lock-in amplifier in combination with a data acquisition module, which records the output signal of the balanced photodetector and the digitized position of the fast-oscillating delay line (shaker). It should be noted that the high repetition rate of 100 kHz allows us to use the fast delay line and retrieve a single THz trace in sub-second time interval. The modulation frequency of the pump beam is used as a reference for the lock-in amplifier, and is set to 2.78 kHz. A bandwidth of 250 Hz is chosen for the low-pass filter of the lock-in amplifier, and the frequency of the shaker to sample the THz trace is set to 0.5 Hz with total travel range of 15 ps (30 ps delay time due to forward/backward path).

Fig. 4 (a) shows the THz trace in the time domain averaged over 300 traces and recorded in purged condition with Nitrogen ($N_2$) gas at a relative humidity of 6%. The corresponding power spectrum on a logarithmic scale is obtained by Fourier transform from the measured THz trace shown in Fig. 4 (b). Another measurement is carried out with the THz beam blocked, shown in dark grey in Fig. 4 (b). This "dark" trace captures the electronic noise floor as well as the influence of beam pointing from imperfect alignment of the shaker, and it is used to obtain more precise determination of the DR in frequency domain. The spectrum has a wide bandwidth that spans more than 9 THz with a peak DR of 55 dB at approximately 1 THz. The smooth, broadband spectrum is facilitated by favorable phase-matching conditions at the 1036 nm driving wavelength and the use of thin MNA as generation and detection crystals. It should be mentioned, that the continuously recorded EOS traces are analyzed and processed in our free open-source software *parrot*, offering advanced correction schemes for highest DR [32,33].

To verify the generated spectral bandwidth, we numerically model the THz generation process in MNA as it was mentioned previously. The blue shaded area in Fig. 4 (b) represents the simulated spectrum as generated in 730 μm MNA (labeled as "sim."). Additionally, we consider the transmission response of the used Ge filter which affects the simulated spectrum as it is shown as dashed blue line (labeled as "sim. × Ge"). It shows an excellent agreement with the measured spectrum at the position of the phase-matching and absorption dips in the spectral range. The small mismatch of the amplitude of higher frequencies can be attributed to the low-pass filter of the lock-in amplifier in combination with the shaker and the filtering characteristics of OAPs.

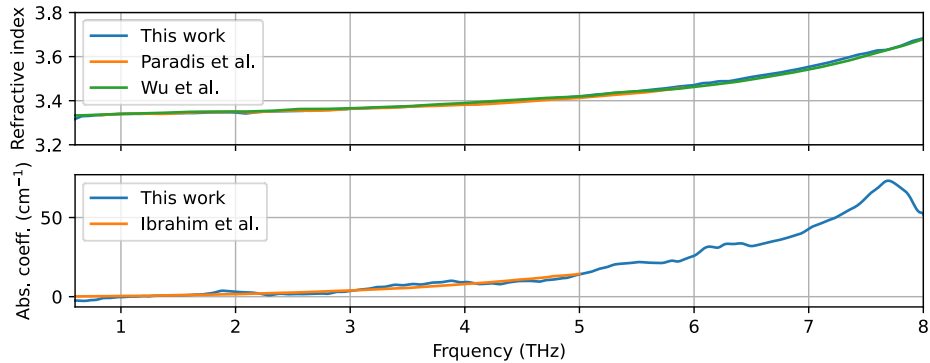

Fig. 5. The refractive index and absorption coefficient of a GaP wafer with thickness of 475 μm placed in the first focus of THz beam. The results are compared to [31], [34] and [35].

In view of future potential applications in nonlinear THz spectroscopy, we provide the spatial image of the THz spot (shown in the inset of Fig. 2 (b)) and the estimate for the THz peak electric field reached at the maximum average power of 11 mW. By using the EOS electric field and the spatial diameter of the THz beam, the estimated THz peak electric field is 212 kV/cm. Therefore, this source can also be attractive for nonlinear spectroscopy.

Although high-power and broadband THz sources based on organic crystals are appealing for a wide variety of applications, the day-to-day reproducibility of the experimental conditions and stability of the THz radiation of these sources has remained poorly explored and often contested in the community. To address these concerns, we perform a long-term measurement to investigate the stability THz generation. The inset in Fig. 4 (a) indicates the maximum amplitude of the EOS trace for 5 days in a row. Each day, a 10-minute measurement is performed with the pump irradiating the crystal. The measurements are done daily in identical condition without any realignment, since the pump laser beam is stabilized using an active laser beam stabilization system. The plot shows the average of the peak-peak amplitude as well as the standard deviation from ensemble of traces belonging to one day in an error bar. The source has an amplitude fluctuation of less than 2% over these 5 days. This result demonstrates a good stability and day-to-day reproducibility.

### 3.2 Characterization of the complex refractive index using THz-TDS

One of the applications of THz-TDS is to determine the optical properties of materials in the frequency range that the THz source can provide. Our source provides a unique tool for characterizing the THz properties of various materials over a wide bandwidth up to 8 THz, demonstrating high DR range performance throughout the entire available spectrum. We note that typical commercial THz-TDS systems can generally only extract THz parameters reliably in the frequency range below 5 THz, due to low DR at higher frequencies. In contrast, the exceptional performance of the all-organic THz-TDS was recently demonstrated in [36] where the complex refractive of an organic crystal OH1 was characterized from 1.5 THz to 12 THz. In another experiment, an organic crystal based-TDS with THz power of 170 nW is used to characterize air (with humidity of 2%) up to 8 THz [21]. The same reference reported a DR of approximately 70 dB in the frequency domain. It should be noted that a background trace is not included, which makes it difficult to assess the true, frequency-resolved DR. Furthermore, it took 2 h to obtain the averaged spectrum due to the low THz average power. The main advantage of our source compared to [21,36] is its significantly higher THz power combined with a high repetition rate. This combination allows for the extraction of material properties, especially for samples with high THz attenuation, in much shorter measurement times (5 minutes in our case).

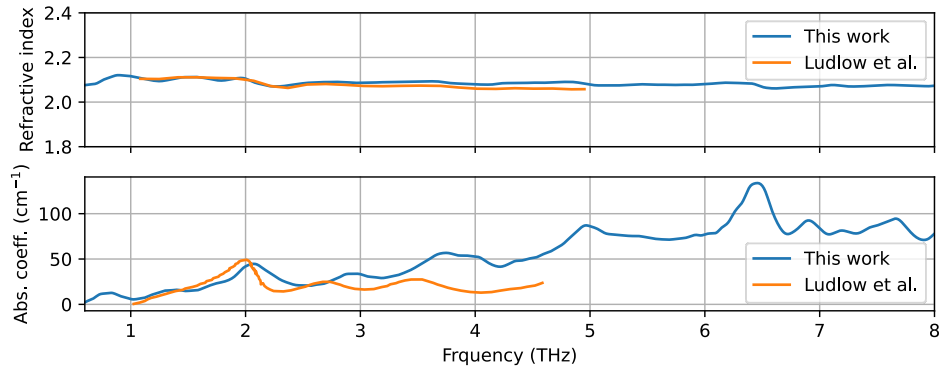

Fig. 6. The refractive index and absorption coefficient of a NMBA crystal with thickness of 490 μm placed in the first focus of THz beam. The results are compared to [37].

To prove the performance of our setup, the complex refractive index of GaP with a thickness of 475 μm and NMBA with a thickness of 490 μm in a broad bandwidth of up to 8 THz is extracted. For the THz transmission measurements, each of the samples is placed in the common focal point of the first and the second OAP. The whole THz path is purged with $N_2$ gas, resulting in a relative humidity of 6%. As is the case with conventional spectroscopic experiments, a complete set of data is made of two measurements: The first measurement is taken after the THz pulse has propagated through the sample, while the second one is performed in the reference medium (in this case $N_2$ gas) without the sample. Additionally, in order to process the THz traces, two dark traces are recorded as well. To extract the refractive index and absorption coefficient, we use *phoeniks* [24]. It should be mentioned that all spectroscopy experiments are conducted with the above-mentioned Ge filter in the THz path which reduces the THz electric field to 85 kV/cm.

The extracted refractive index and absorption coefficient of GaP in the frequency range up to 8 THz is shown in Fig. 5. To confirm the validity of the reconstructed refractive index, we compare these results with the calculated results in [31] and measured results from [34] which shows an excellent agreement between calculated and measured refractive index. Our measured results for the absorption coefficient are compared with those from of [35], also showing good agreement.

The second sample to extract the complex refractive index is the organic crystal NMBA with a thickness of 490 μm. The results are plotted in Fig. 6. We present a much wider frequency range compared to the previous work reported in [37]. The real part of refractive index is compared to the values reported in [37], and they are in an excellent agreement. The THz absorption coefficient of NMBA reaches a maximum value of approximately 100 cm$^{-1}$ at 6.4 THz. A comparison of the present work results with those of reference [37] indicates that up to 3.5 THz, the results are in good agreement. However, beyond this point, the results begin to slightly diverge from each other. This might be due to the difference in the crystal roughness and scattering loss of the NMBA surface. As the higher THz frequency components with smaller focus scatter by the crystal roughness more than the lower frequencies with a larger focus, the mismatch in the absorption coefficient increases. Additionally, a small misalignment of the NMBA generation axis with respect to the THz polarization can be results a difference in the absorption coefficient values for both cases.

This developed THz-TDS excels in providing high THz average power at higher repetition rates (> 100 Hz), which significantly reduces the measurement time for tasks such as extracting the complex refractive index. Although state-of-the-art organic crystal-based sources have shown impressive results in terms of bandwidth and electric field, they often lack the THz average power needed to achieve DR and high bandwidth results in a short amount of measurement time.

## 4. Conclusion and outlook

In conclusion, we demonstrate a high power and broadband, table-top THz-TDS based on OR in the organic crystal MNA. In optimized conditions, using 8 W of driving power (after chopping) on the crystal, we reach 11 mW of THz average power with a broad bandwidth extending up to 9 THz, at a peak DR of 55 dB. The use of MNA also as a broadband detector allows us to maintain a high DR > 40 dB up to 8 THz. To the best of our knowledge this is the first report of a THz source based on MNA using 1030 nm high average power excitation in combination with an MNA based detection. Our TDS provides a unique combination of usable bandwidth, DR and average power for a wide range of applications. We used this setup to characterize the THz properties of GaP and NMBA (another organic crystal from the same family) over a bandwidth > 8 THz, exemplifying the high DR performance over the wide bandwidth. In addition, we examine the stability of the THz source by measurements at subsequent days. Furthermore, our source achieves peak electric field > 200 kV/cm, making it an excellent candidate for nonlinear spectroscopy. We believe the TDS demonstrated – based on commercially available high-power laser technology- offers great promise for broadband THz science and technology.

**Funding.** This work was supported in part by DFG of the SFB/TRR196 MARIE project M01 and C07 and the DFG project PR1413/3–2, in part by the DFG under Germany's Excellence Strategy—EXC-2033— Projektnummer 390677874—Resolv, and in part by project "terahertz.NRW" program "Netzwerke 2021," an initiative of the Ministry of Culture and Science of the State of Northrhine Westphalia. We acknowledge support by the DFG Open Access Publication Funds of the Ruhr-Universität Bochum.

**Disclosures.** DJM and JAJ disclose that they are co-founders of Terahertz Innovations LLC.

**Data Availability.** Data underlying the results presented in this paper are not publicly available at this time but may be obtained from the authors upon reasonable request.